# On the Limits of Information Retrieval in Quantum Mechanics

P. B. Lerner[1]

Abstract

The widely considered assertion is that the unitarity of quantum mechanical evolution assures the "preservation" of information. It is even promoted in popular literature as an established fact. (Susskind, 2008) Yet, a simple chain of reasoning demonstrates that: 1) almost any evolutionary operator can be well approximated by a degenerate (finite-rank) operator and 2) one needs an eternity to retrieve information exactly from a non-stationary quantum state and to distinguish between arbitrary unitary operator and its finite-dimensional approximations.

**Introduction**

The preservation of quantum mechanical information as a result of unitarity of evolution operators—usually assumed in the standard schemes of quantum mechanics—acquired the status of popular mythology, which is repeated in numerous literature sources without much debate. It is claimed to be a law of nature akin to the conservation of energy or growth of entropy. (Susskind, 2008) In this paper, I re-capitulate two elementary facts. First, unitary evolution operators have many degenerate (finite-rank) operators in their immediate vicinity. Second, checking that a given operator is unitary within the confines of paradigmatic quantum mechanics is, in general, impossible.

One of the first demonstrations of these, quite routine conclusions from the general formalism of quantum mechanics in the context of irreversibility of quantum mechanical experiments belongs to genial and untimely deceased Asher Peres. (Peres, 1984) In a simple (now available to any undergraduate student) and elegant numerical experiment, he proved that if there is an uncertainty in determination of a Hamiltonian of the quantum mechanical system, the "average" long-term evolution of the <u>family</u> of indistinguishable QM systems is irreversible.

---

1 Peter B. Lerner, Scitech Associates, LLC, Ithaca, NY. pblerner@syr.edu, pbl2@psu.edu

This paper, in its first part, follows a different approach. Within the confines of what I call "primitive" quantum mechanics—a formalism usually described in textbooks as The Quantum Mechanics—I demonstrate that there exists many finite-dimensional approximations of the original Hilbert state, which produce the results of measurement of any observable, which are arbitrarily close to the true quantum mechanical expectations. This is a direct consequence of well-known approximation property in Banach spaces. Because, the extension of the approximation of the original unitary evolution operator on a finite-dimensional space can have infinite-dimensional ("fat") kernel on the original Hilbert space, this proves that there are many degenerate operators in the arbitrary vicinity of the true QM evolution.

In the second part, I use considerations analogous to the Mandelstam-Tamm uncertainty relation to demonstrate that an infinite precision for the quantum-mechanical measurement, in general, requires infinite time of measurement. For a physical experiment, limited in time, i.e. having a beginning and an end, there is no general possibility to distinguish between a measurement made by a perfect, or a "coarse-grained" device. Because of that, reconstruction of an original quantum state back to its origins is usually impossible and the unitary nature of quantum-mechanical evolution is irrelevant. Presented results are not the only limits on "computability" in quantum mechanics and the author hopes to present other limitations in subsequent publications.

1.          **Finite-dimensional Approximation of Quantum Observables**

A. Primitive quantum mechanics

In this section, we describe the structure of "primitive" (or standard) quantum mechanics. The author acknowledges caveats, both theoretical and practical, in this description but this is what is typically meant by "quantum mechanics" in conventional textbook treatments. (see e.g. Chapter 2 of Nielsen and Chuang, 2001 with the reference to Sakurai, 1995 and Cohen-Tannoudji, Diu and Laloe, 1977) Namely, given is a separable Hilbert space over the complex numbers field $L$, the vectors $\Psi \in L$ of which are called states. The select Hermitian injection operator $\hat{H}: L \to L$ is called Hamiltonian. There are other Hermitian operators $\hat{A}$ on $L$, which are called observables. All linear operators on the Hilbert space constitute an algebra over the field of complex numbers. All bounded operators on $L$:

$$\langle\Psi|B|\Psi\rangle < \infty$$

have a natural norm defined by the identity:

$$\|B\| = \sup_{\langle\Psi|\Psi\rangle=1} \frac{\langle\Psi|B|\Psi\rangle}{\langle\Psi|\Psi\rangle} \qquad (1)$$

This norm, under some generic assumptions, turns algebra of bounded operators on *L* into a Banach space.

Quantum mechanical evolution is described by the exponential of the Hamiltonian:

$$U = e^{i\hat{H}t} \qquad (2)$$

as transformations of the state vector. This exponential, under certain mild conditions is a unitary operator.[2] (Dunford and Schwartz, 1971, Chapter VIII) Note, that, first, we consider explicitly relativistically non-covariant formulation of the quantum mechanics. Second, our Hamiltonian is time-independent to simplify the analytic reasoning. Otherwise, we have to deal with time ordering of exponential operators. Comment on these limitations will be provided in Part II of the present work.

B. Convergence of truncated observables

First, we note the following obvious property of any primitive quantum mechanics as a system. Separable Hilbert space *L* possesses a complete system of mutually orthogonal eigenvectors. (Dunford and Schwartz, 1971) Because we admitted the existence of a Hamiltonian operator, rotation to one of these bases turns the energy norm into a diagonal Dirichlet form.

For a mathematically inclined reader, the rest of the Section 1B is a reminder of the approximating property for the Banach space of observables. (Kato, 1966, Chapter IX, Lindenstrauss and Tzafiri, 1977, Fabian, 2001, Chapter 15) However, the approximating property in conventional context is proven (or rejected) for the case of operators in general Banach spaces. For us, it is sufficient to deal with a much more limited situation of non-compact linear operators (observables) on a Hilbert space.

The truncation of this system to the first n vectors defines a projector $P_n$ and orthogonal projector $Q_{n>N}$ on the Hilbert subspace orthogonal to the first *n* vectors. Obviously, for each *n*:

---
[2] In particular, operator *H* is not necessarily required to be bounded. (Derezinsky, 2013)

$$Det_{n'}(P_n B P'_n) = 0 \tag{3}$$

for an arbitrary non-singular matrix $B$ on a finite-dimensional Hilbert subspace with dimension $n' > n$ engendered by the same sequence of projection operators.

Now we can formulate the first lemma.

**Lemma 1**

For any $\varepsilon > 0$, ∃ $\exists n > 0$, such that the expectation of an arbitrary observable obeys at all times:

$$\frac{|\langle \Psi | \hat{A}(t) - P_n \hat{A}(t) P'_n | \Psi \rangle|}{|\langle \Psi | \Psi \rangle|} < \varepsilon \tag{4}$$

This lemma is an automatic consequence of the approximation property of the Banach spaces. Yet, I emphasize demonstration more limited in scope but which is applicable when observables are unbounded operators.

**Corollary 1 from Lemma 1**

For any $\varepsilon > 0$ $\exists n' > n$ such that $\hat{U}_n = P_n U P'_n$, $|Det_{n'} \hat{U}_n| = 0$ and $U$ from Equation (2):

$$\frac{|\langle \Psi | U - \hat{U}_n | \Psi \rangle|}{\langle \Psi | \Psi \rangle} < \varepsilon \tag{6}$$

Above, the $Det_{n'} \hat{U}_n$ is the determinant of the operator $U_n$ acting on the $n'$-dimensional subspace of the original Hilbert space $L$. The Corollary 1 follows from taking $\hat{A}_0 = \hat{I}$ as an observable in Lemma 1. Note that this operator is not compact. Corollary 1 demonstrates that for any unitary operator of the form of Equation (2), there is an operator with non-zero (in fact, infinite-dimensional kernel) which is arbitrarily close to the original $U$ by the "natural" operator norm of the Banach space.

C. Construction of the measurement device

Here we shall quantify a "primitive" theory of measurement, which is a condensed version of a standard textbook theory of measurement (Chuang and Nielsen, 2001) and is close to the axiomatic formulation by Mackey (1960).

Namely, a measurement device is a non-empty <u>family</u> of (weakly convex) measures $\mu_\alpha$ on a (locally) compact subalgebra $\bar{C}$ of the algebra of the observables, obeying a conventional system of axioms for measure. (McKay, 1960) Then, the measurement operator for an observable $\hat{A}(t)$ is defined by:

$$\hat{M}_\alpha = \int \hat{A}(t) d\mu_\alpha \qquad (7)$$

wherever this integral exists. Note that even if the integral of Equation (7) is well-defined for *t=0*, it is not necessary that it exists for all *t>0*. Thus, it makes sense to define a subalgebra $\kappa_t \subseteq \bar{C}$, for which Equation (7) is well-defined for a given measurement device. The probability distribution $P_\alpha$ for the measurement results is described by a conventional Born formula:

$$P_\alpha(t) = \int_{\kappa_t} \langle \Psi | \hat{A}(t) | \Psi \rangle d\mu_\alpha \triangleq P_\alpha(t|\kappa_t) \qquad (8)$$

The example of the measurement device is easy to define. Namely, for each finite-dimensional family of the state vectors, which can also be identified with functionals on the original Hilbert space *L* through conventional duality, $\{\Psi_i\}$, *i=1, N*, a measurement device can be described by the following Equation (9):

$$\mu_\alpha(t) = \sum_i \rho_{\alpha i}(t) |\Psi_i\rangle\langle\Psi_i| \qquad (9)$$

Obviously, from the normalization condition:

$$\sum_\alpha \rho_{\alpha,i}(t) = 1 \qquad (10)$$

In 1957, Gleason proved a remarkable theorem that the expressions of Equation (9), if we allow the index to span all basis vectors, under certain seemingly innocuous conditions exhaust all the possible measures (Gleason, 1957, Mackey, 1960). Because of that, the measures of the type (9) can be identified with all convex measures on finite-dimensional subspaces of *L*.

In the parlance of Mackey (1960, Chapter 2), any measurement device can be decomposed into (or, more, colloquially, *answers* to) a finite series of *questions*, which Mackey defined as projectors into one-dimensional subspaces of an original Hilbert space $L$.[3]

---

3   The name of these projection operators as questions is justified by their eigenvalues, which can be only 0 or 1.

This formulation of measurement theory agrees with an intuitive notion that a "sensible" measurement device has a finite number of internal states subject to experimental monitoring. Then, the (optional) time dependence of the coefficients $\rho_{\alpha,i}(t)$ can be naively interpreted as the values of "dials" changed by the experimentalist in course of the measurement. The monitoring itself, however, can be accomplished at an infinite number of points—or even be continuous—in some contexts it might be beneficial to identify parameter α with the reading *T* of an internal clock of the device.

In the above formulation, there is no separate macroscopic measurement device obeying "classical laws" and the observed particle obeying quantum mechanics. "Macroscopic" simply means that the measurement device has one or more degrees of freedom, which are impossible or impractical to control quantum mechanically and for which one must use statistical description.

## 2. Time Limits on the Distinguishability of the Quantum Observables

A possibility to approximate any Hermitian operator on a Hilbert state by a weakly compact (i.e. irreversible) operator does not mean, in general, that one cannot make measurements with a complete preservation of the initial observable. Yet, in this section, we demonstrate that it, in general, is incredibly difficult. In particular, we point out that more and more precise measurements have diminishing returns in terms of the Fisher information. (Bickel and Doksum, 2000, Fröwis, 2012)

This fact means that one cannot, in general, improve the resolution of quantum mechanical observables indefinitely, though this possibility may exist in very special cases.

A. Criterion for the efficiency of quantum measurement

The existence of many similar but less informative quantum states in the vicinity of the "true" state, which is the product of unitary evolution of an original one, does not necessarily mean the impossibility of distinguishing between them. However, to accomplish this distinction, one needs at least a theoretical opportunity to increase the precision of a quantum measurement procedure indefinitely. To investigate this matter, we need to elaborate further on the protocol of quantum measurement.

Every quantum mechanical measurement relies on several idealized steps. First, the measurement apparatus is prepared in a quantum certain state at *t=0*. Then, it is put in contact with the system undergoing the measurement (which is conventionally called particle) and the probabilities of transition of the measurement apparatus into its internal states are read out at some time $t>0$.

To describe this situation, we need to distinguish between the Hamiltonian of the system undergoing the measurement $H_S$ and the Hamiltonian of the measurement apparatus-system interaction $V$. In this Section, we discuss the criteria for the quality of the quantum measurement. Let the total wavefunction be as follows:

$$|\Psi(t)\rangle = \sum_{i=1}^{N} c_i |\psi_i(t)\rangle + \sum_{j=N+1}^{\infty} d_j |\psi_j(t)\rangle \equiv P_N |\Psi(t)\rangle + Q_N |\Psi(t)\rangle \tag{11}$$

or

$$|\Psi(t)\rangle = \sum_{i=1}^{N} c_i e^{iE_i t/\hbar} e^{i\hat{V}t/\hbar} |\psi_i(0)\rangle + \sum_{j=N+1}^{\infty} d_j e^{iE_j t/\hbar} e^{i\hat{V}t/\hbar} |\psi_j(0)\rangle \tag{12}$$

Here, the index $i$ refers to the (finite) number of eigenstates of the measurement apparatus, which are brought into contact with a particle, and the index $j$—to the potentially infinite number of eigenstates of the measured particle. $P_N$ and $Q_N$ are the orthogonal projectors on the Hilbert subspaces formed by the eigenstates of the unperturbed Hamiltonian. Interaction $V$ is considered as perturbation.

Using generalized Born rule for the transition probability, we obtain:

$$\begin{aligned}P_\alpha(t) = &\sum_i |c_i|^2 \rho_{\alpha i}|\langle\psi_i(0)|e^{i\hat{V}t/\hbar}|\psi_i(0)\rangle|^2 + \sum_{i,j} c_i \bar{d}_j e^{i(E_i-E_j)t/\hbar} \rho_{\alpha i}|\langle\psi_i(0)|e^{i\hat{V}t/\hbar}|\psi_j(0)\rangle|^2 \\ &+ \sum_{i,j} \bar{c}_i d_j e^{-i(E_i-E_j)t/\hbar} \rho_{\alpha i}|\langle\psi_j(0)|e^{i\hat{V}t/\hbar}|\psi_i(0)\rangle|^2 + \sum_i |d_j|^2 \rho_{\alpha i}|\langle\psi_j(0)|e^{i\hat{V}t/\hbar}|\psi_j(0)\rangle|^2\end{aligned} \tag{13}$$

If the measurement apparatus and the measured system were never brought in contact with each other, the internal states of measurement apparatus also would evolve but according to the formula:

$$P'_\alpha(t) = \sum_i |c_i|^2 \rho_{\alpha i}|\langle\psi_i(0)|e^{i\hat{V}t/\hbar}|\psi_i(0)\rangle|^2 + \sum_i |d_j|^2 \rho_{\alpha i}|\langle\psi_j(0)|e^{i\hat{V}t/\hbar}|\psi_j(0)\rangle|^2 \tag{14}$$

A natural criterion of the efficiency (quality) of our measurement would be the analog of the Kolmogorov-Smirnov distance (Shorack and Wellner, 2009):

$$Q = \max_t |P_\alpha(t) - P'_\alpha(t)| \tag{15}$$

This criterion identifies the quality of measurement with the K-S distance between the distributions of the probability of internal states of the measurement device in cases of free evolution and the evolution affected by the interaction with a particle.

The quality of measurement is equal to:

$$Q = \max_t [.]$$
$$\left| \sum_{i,j} c_i \bar{d}_j e^{i(E_i-E_j)t/\hbar} \rho_{\alpha i} |\langle \psi_i(0)|e^{i\hat{V}t/\hbar}|\psi_j(0)\rangle|^2 + \sum_{i,j} \bar{c}_i d_j e^{-i(E_i-E_j)t/\hbar} \rho_{\alpha i} |\langle \psi_j(0)|e^{i\hat{V}t/\hbar}|\psi_i(0)\rangle|^2 \right| \tag{16}$$

below we shall make a useful notation $\gamma_{ji} = \bar{\gamma}_{ij} = e^{-i(E_i-E_j)t/\hbar}$. With it,

$$Q = \max_t \left| \sum_{i,j} c_i \bar{d}_j \gamma_{ij} \rho_{\alpha i} |\langle \psi_i(0)|e^{i\hat{V}t/\hbar}|\psi_j(0)\rangle|^2 + \sum_{i,j} \bar{c}_i d_j \gamma_{ji} \rho_{\alpha i} |\langle \psi_j(0)|e^{i\hat{V}t/\hbar}|\psi_i(0)\rangle|^2 \right| \tag{17}$$

We shall use this expression in the next section to prove that an arbitrarily accurate quantum measurement might take arbitrarily long time.

B. Temporal limits on the resolution of the quantum devices

In this section, we prove that the resolution of the quantum devices cannot be, in general, increased indefinitely if the measurement lasts a finite time. This limitation has an intimate connection with the Mandelstam-Tamm uncertainty relation. (Mandelstam, Tamm, 1945)

We compare the quality of the measurement achievable by a "perfect" measurement device and a "coarsened" measurement device, in which the interaction Hamiltonian is truncated to a finite number of states. The quality of the measurement made by a perfect device is expressed by the Equation (18), in which the summation indexes are made explicit:

$$Q = \max_t [|q(t)|] = \max_t [.]$$
$$\left[ \left| \sum_{i=1, j=N+1}^{i=N, j=\infty} c_i \bar{d}_j \gamma_{ij} \rho_{\alpha i} |\langle \psi_i(0)|e^{\frac{i\hat{V}t}{\hbar}}|\psi_j(0)\rangle|^2 + \sum_{i=1, j=N+1}^{i=N, j=\infty} \bar{c}_i d_j \gamma_{ji} \rho_{\alpha i} |\langle \psi_i(0)|e^{\frac{i\hat{V}t}{\hbar}}|\psi_j(0)\rangle|^2 \right| \right] \tag{18}$$

The quality of measurement by the coarse device is expressed by a similar formula with different limits

$$\tilde{Q} = \max_t = \max_t [|\tilde{q}(t)|]$$

$$\left[ \left| \sum_{i=1, j=N+1}^{i=N, j=N_1} c_i \bar{d}_j \gamma_{ij} \rho_{\alpha i} |\langle \psi_i(0)| e^{\frac{i\hat{V}t}{\hbar}} |\psi_j(0)\rangle|^2 + \sum_{i=1, j=N+1}^{i=N, j=N_1} \bar{c}_i d_j \gamma_{ji} \rho_{\alpha i} |\langle \psi_i(0)| e^{\frac{i\hat{V}t}{\hbar}} |\psi_j(0)\rangle|^2 \right| \right]$$
(19)

By construction, $Q \geq \tilde{Q} \geq 0$ .

According to the triangle inequality:

$$0 \leq F \equiv |Q| - |\tilde{Q}| \leq \max_t (|q(t) - \tilde{q}(t)|) .$$
(20)

It can be demonstrated that, in general, the fidelity factor of the measurement $F$, i.e. the difference of readout for a perfect and coarse devices is infinitesimal for any finite time of the measurement procedure.

**Theorem 1**

For any $\varepsilon > 0$ and sufficiently low correlations between the elements of the matrix $M_{ij} = c_i \bar{d}_j \gamma_{ij}$ :

$$\sum_{j=N_1}^{\infty} \Re[M_{ij}], \Im[M_{ij}] \xrightarrow{N_1 \to \infty} 0$$
(21)

there exists a numerical constant A such that

$$F \leq A t \varepsilon$$
(22)

The condition of Equation (21) on the stochasticity of the matrix elements $M_{ij}$ is natural to be called "Peres condition" after Asher Peres, who was first to connect non-integrability of the perturbation Hamiltonian with the irreversibility in quantum mechanics (Peres, 1984). This concludes our reasoning that experimental distinction between correct, i.e. information-preserving original state and its finite-dimensional approximation is a virtual impossibility. The above caveat is essential because our demonstration is provided in restrictive conditions, which can sometimes be violated. Heuristically, Theorem 1 states a rather simple fact: for a general measurement procedure adding more and more eigenstates brings diminishing returns to the Fisher information (Fröwis, 2012) obtained through that measurement. The proof requires a technical Lemma 2 (Appendix, see also Schmudgen, 2012 and Derezinski, 2013).

**Lemma 2**

The evolution operator of Equation (2) is a smooth function of t in the complex band -∞<Re(t)<∞, 0≤Im(t)<u<∞.

Consequence of the proof of Theorem 1 is the following inequality. (Equation (A.10) of the Appendix) Because of the convergence of the series of matrix elements, for a large enough $N_I$, the infinite sum $\sum_{N_1}^{\infty} \tilde{V}_{ij}$ converges to a finite limit. Then,

$$f_i \leq 4 K \rho_{\alpha,i} \left| \sum_{j=N_1}^{\infty} \Re[M_{ij}]\left(\frac{\tilde{V}_{ij} t}{\hbar}\right) \right| \leq C \varepsilon \frac{t}{\hbar} \left| \sum_{j=N_1}^{\infty} \tilde{V}_{ij} \right| \leq \tilde{C} \frac{t}{\hbar} \varepsilon \qquad (23)$$

Because the number of internal states of the measurement device is finite, so is *F*—fidelity of distinction between precise and coarse devices:

$$F < A t \varepsilon \qquad (23')$$

In Equation (23), $A, C, \tilde{C}, K < \infty$ are numerical constants. Due to the Equation (22-22'), the quality of measurement is infinitesimal for any finite time. Note that our set of inequalities breaks down for interaction-dependent characteristic times defined by:

$$t > t_0 = \frac{\pi}{2} \frac{\hbar}{\max_{i,N_1}\left(\tilde{V}_{i,N_1}\right)} \qquad (24)$$

because for the times longer than $t_0$, the convexity of the sin(x) breaks down. The result is that, in general, one cannot make a distinction between "coarse" and "perfect" measurement device during a finite time of arbitrary experiment. Together with the conclusions of the Section 1, we deduce that the distinction between unitary operator on a Hilbert space and its finite-dimensional approximation (having infinite-dimensional kernel) is impossible.

**Conclusion**

We proved, albeit in a very simplified and restrictive framework, that within the confines of non-relativistic quantum mechanics, the distinction between a unitary operator and its finite-dimensional approximations cannot, in general, be observable. Relativistic framework intuitively seems only to complicate matters with the observability. However, it seems to this author that, at least, the demonstration where $t$ is not a parameter but a coordinate on the future space-time cone can be instructive.

While the information is preserved during unitary quantum-mechanical evolution, in general there is no experimental possibility to reconstruct an original unitary operator from the values of any observable during a finite time of measurement. The loss of information about initial quantum state is inevitable.

Yet, the existence of symbolic Peres condition (stochastic distribution of eigenvalues, Section 2B) demonstrates that this conclusion is not universal and in some, well-designed experiments, complete reconstruction of an original unitary operator and backdating of a quantum state to its origin may be possible. Whether it is possible with all naturally occurring astrophysical systems is another question entirely.

Appendix. **Proofs of the Lemma 1, Theorem 1 and Lemma 2**

Proof of Lemma 1

First, note that for each n, $\langle \Psi P'_n | P_n \Psi \rangle < \infty$ and $\langle \Psi Q'_n | Q_n \Psi \rangle < \infty$. I.e. for each $\delta > 0$, $\exists n$, such that $|\sum_{n'>n} c*_{n'} c_{n'}| < \delta$ where $c_{n'} = \langle \Psi | P_{n'} \Psi \rangle$. This follows from the convergence of the eigenfunction expansion. Second, for each $\varepsilon > 0$ $\exists n$, such that $|\langle \Psi | Q_n | \Psi \rangle| < \varepsilon$.

Given that

$$\hat{A}(t) = e^{-i\hat{H}t} A_0 e^{i\hat{H}t} \qquad (A.1)$$

the denominator of the Equation (4) of the main text :

$$|\langle \Psi | \hat{A}(t) - P_n \hat{A}(t) P'_n | \Psi \rangle| = |\langle \Psi | (P_n + Q_n) \hat{A}(t)(P_n + Q_n)' - P_n \hat{A}(t) P'_n | \Psi \rangle|$$
$$\leq |\langle \Psi | Q_n \hat{A}(t) Q_n' | \Psi \rangle| + |\langle \Psi | Q_n \hat{A}(t) P_n' + P_n \hat{A}(t) Q_n' | \Psi \rangle| + |\langle \Psi | e^{i\hat{H}t} [P_n, A_0]^c e^{-i\hat{H}t} | \Psi \rangle| \qquad (A.2)$$

where $[P_n, A_0]^c = P'_n A_0 P_n - P_n A_0 P'_n$. Then it is sufficient to chose $\delta < \varepsilon |\langle \Psi | \Psi \rangle| / 2$ and n such that $|\langle \Psi | Q_n | \Psi \rangle| < \varepsilon |\langle \Psi | \Psi \rangle| / 2 |\langle \Psi | \hat{A}(t) | \Psi \rangle|$. The latter is always possible because states have a finite norm. The cross terms with $P_n$ and $Q_n$ are eliminated through orthogonality of the projectors.

Proof of Lemma 2

Following Schmudgen (2012), we assume that the interaction operator obeys the inequality for each $x \in L$ where A and B are numerical constants:
$$\langle x, \hat{H} x \rangle \geq A \langle x, \hat{V} x \rangle + B \qquad (A.3)$$
For the interaction operator, there exists a spectral representation (VIII.1-2, Danford and Schwartz, 1971):
$$e^{i\hat{V}t} = \int e^{-\lambda t} (\lambda I - \hat{V})^{-1} d\lambda \qquad (A.4)$$
This representation can be differentiated in *t* any number of times without affecting its convergence. If the integral (A.4) exists then there is $\lambda_0$ such that:

$$\|\int_{-\infty}^{-\lambda_0} e^{-\lambda t}(\hat{V}-i\lambda I)^{-1} d\lambda\|+\|\int_{\lambda_0}^{\infty} e^{-\lambda t}(\hat{V}-i\lambda I)^{-1} d\lambda\|<\varepsilon \qquad (A.5)$$

For the n-th derivative, $\lambda_0$ can be replaced by $\lambda_0 + 2n \log \lambda_0$ with the preservation of the same inequality. Jourdan's lemma assures the convergence of (A.4) in the complex band.

Proof of Theorem 1

$$F \equiv \sum_{i=1}^{N} f_i \leq \max_t [.]$$
$$\left[ \left| \sum_{i=1, j=N_1}^{i=N, j=\infty} c_i \bar{d}_j \gamma_{ij} \rho_{\alpha i} |\langle \psi_i(0)| e^{\frac{i\hat{V}t}{\hbar}} |\psi_j(0)\rangle|^2 + \sum_{i=1, j=N_1}^{i=N, j=\infty} \bar{c}_i d_j \gamma_{ji} \rho_{\alpha i} |\langle \psi_i(0)| e^{\frac{i\hat{V}t}{\hbar}} |\psi_j(0)\rangle|^2 \right| \right] \qquad (A.6)$$

Separating, as usual, Hermitian and anti-Hermitian part of the interaction evolution operators (Loudon, 2000)

$$e^{i\hat{V}t/\hbar} = \hat{\cos}(\frac{\hat{V}t}{\hbar}) + i\hat{\sin}(\frac{\hat{V}t}{\hbar}) \qquad (A.7)$$

we obtain an equality:

$$f_i \leq \left| \sum_{j=N_1}^{j=\infty} c_i \bar{d}_j \gamma_{ij} \rho_{\alpha i} \langle \psi_i(0)|\hat{\sin}(\frac{\hat{V}t}{\hbar})|\psi_j(0)\rangle + \sum_{j=N_1}^{j=\infty} \bar{c}_i d_j \gamma_{ji} \rho_{\alpha i} \langle \psi_i(0)|\hat{\sin}(\frac{\hat{V}t}{\hbar})|\psi_j(0)\rangle \right| \qquad (A.8)$$

where we have used the operator identity $\hat{\cos}^2 \hat{X} + \hat{\sin}^2 \hat{X} = \hat{1}$ and the mutual orthogonality of the wave function systems of the apparatus $\{\psi_i(0)\}$, $i=1, N$ and the particle $\{\psi_j(0)\}$, $j=N+1, \infty$. This situation will be discussed below.

Because of the convexity of the sin(x) function in the first quadrant $0 \leq x \leq \pi$, for sufficiently small *t*, we can apply Mond-Pečaric theorem (Theorem 91, p. 62, Dragomir (2011)). The theorem places limitations on the convex functions of operators with the spectrum, which is concentrated on the closed interval on the real axis. The inequality in (A.8) can be replaced by the inequality containing the absolute values of matrix elements $V_{ij>N_1}$ :

$$K\left|\sum_{j=N_1}^{j=\infty} M_{ij}\rho_{\alpha i}\sin\left(\frac{2|\langle\psi_i(0)|\hat{V}|\psi_j(0)\rangle|t}{\hbar}\right)+\sum_{j=N_1}^{j=\infty}\bar{M}_{ij}\rho_{\alpha i}\sin\left(\frac{2|\langle\psi_i(0)|\hat{V}|\psi_j(0)\rangle|t}{\hbar}\right)\right|\overset{f_i\leqslant\lambda}{\leqslant\lambda} \quad (A.9)$$

$$4K\rho_{\alpha,i}\sum_{j=N_1}^{\infty}\Re[M_{ij}]\sin\left(\frac{2\tilde{V}_{ij}t}{\hbar}\right)\leqslant 8K\rho_{\alpha,i}\sum_{j=N_1}^{\infty}\Re[M_{ij}]\left(\frac{\tilde{V}_{ij}t}{\hbar}\right)$$

where $M_{ij}=c_i\bar{d}_j\gamma_{ij}$ and $\tilde{V}_{ij}=|\langle\psi_i(0)|\hat{V}|\psi_j(0)\rangle|$ and K<∞.

Also, from the Lemma2 we know that the evolution operator is a smooth function of t in the complex band -∞<Re(t)<∞, 0≤Im(t)<u<∞. Because of the convergence of the series of matrix elements, for a large enough $N_1$, the infinite sum $\sum_{N_1}^{\infty}\tilde{V}_{ij}$ converges to a finite limit. Then, the Peres condition on the matrix elements $M_{ij}$ assures convergence of the fidelity in the channel *i* to zero:

$$f_i\leqslant 8K\rho_{\alpha,i}\left|\sum_{j=N_1}^{\infty}\Re[M_{ij}]\left(\frac{\tilde{V}_{ij}t}{\hbar}\right)\right|\leqslant C\varepsilon\frac{t}{\hbar}\left|\sum_{j=N_1}^{\infty}\tilde{V}_{ij}\right|\leqslant\tilde{C}\frac{t}{\hbar}\varepsilon \quad (A.10)$$

In Equation (A.10), $C,\tilde{C}<\infty$ is a numerical constant. Due to the Equation (A.10) and finite number of the channels in any measurement device, quality of measurement is infinitesimal for any finite time. Note that our set of inequalities breaks down for characteristic times defined by:

$$t>t_0=\frac{\pi}{2}\frac{\hbar}{\max_{i,N_1}(\tilde{V}_{i,N_1})} \quad (A.11)$$

because for the times longer than $t_0$, the convexity of the sin(x) breaks down.